\documentclass{iaus}

\usepackage{graphicx}
\usepackage{epsf}

\newcommand{\etall}{{et al.~}}

\newcommand{\gta}{\gtrsim}

\newcommand{\masyr}{\>{\rm mas}\,{\rm yr}^{-1}}
\newcommand{\kms}{\>{\rm km}\,{\rm s}^{-1}}

\newcommand{\kpc}{\>{\rm kpc}}

\newcommand{\Msun}{\>{\rm M_{\odot}}}

\newcommand{\Gyr}{\>{\rm Gyr}}

\newif\ifsubmode
\submodefalse

\newif\ifprintfig
\printfigtrue

\title[Magellanic System Kinematics]{Kinematical Structure\\
of the Magellanic System}

\author[van der Marel, Kallivayalil, \& Besla]{Roeland P. van der Marel$^1$,
Nitya Kallivayalil$^2$, \& Gurtina Besla$^3$}

\affiliation{$^1$Space Telescope Science Institute, 3700 San Martin
Drive, Baltimore, MD 21218\\
$^2$MIT, Kavli Inst.~for Astrophysics \& Space Research,
70 Vassar Street, Cambridge, MA 02139\\
$^3$Harvard-Smithsonian Center for Astrophysics, 60 Garden Street,
Cambridge, MA 02138}

\pubyear{2008}
\volume{256}  %% insert here IAU Symposium No.
\jname{The Magellanic System: Stars, Gas, and Galaxies}
\editors{Jacco Th. van Loon \& Joana M. Oliveira, eds.}

\begin{document}

\maketitle

\begin{abstract}
We review our understanding of the kinematics of the LMC and the SMC,
and their orbit around the Milky Way. The line-of-sight velocity
fields of both the LMC and SMC have been mapped with high accuracy
using thousands of discrete traces, as well as HI gas. The LMC is a
rotating disk for which the viewing angles have been well-established
using various methods. The disk is elliptical in its disk plane. The
disk thickness varies depending on the tracer population, with
$V/\sigma$ ranging from $\sim 2$--10 from the oldest to the youngest
population. For the SMC, the old stellar population resides in a
spheroidal distribution with considerable line-of-sight depth and low
$V/\sigma$. Young stars and HI gas reside in a more irregular rotating
disk. Mass estimates based on the kinematics indicate that each Cloud
is embedded in a dark halo. Proper motion measurements with HST show
that both galaxies move significantly more rapidly around the Milky
Way than previously believed. This indicates that for a canonical
$10^{12} \Msun$ Milky Way the Clouds are only passing by us for the
first time. Although a higher Milky Way mass yields a bound orbit,
this orbit is still very different from what has been previously
assumed in models of the Magellanic Stream. Hence, much of our
understanding of the history of the Magellanic System and the
formation of the Magellanic Stream may need to be revised. The
accuracy of the proper motion data is insufficient to say whether or
not the LMC and SMC are bound to each other, but bound orbits do exist
within the proper motion error ellipse.\looseness=-2
\end{abstract}

% if document starts with a section,

\firstsection 

\section{Introduction}
\label{s:intro}

The Magellanic Clouds are two of the closest galaxies to the Milky
Way, with the Large Magellanic Cloud (LMC) at a distance of $\sim 50
\kpc$ and the Small Magellanic Cloud (SMC) at $\sim 62 \kpc$. Because
of their proximity, they are two of the best-studied galaxies in the
Universe. As such, they are a benchmark for studies on various topics,
including stellar populations and the interstellar medium,
microlensing by dark objects, and the cosmological distance scale. As
nearby companions of the Milky Way with significant signs of mutual
interaction, they have also been taken as examples of hierarchical
structure formation in the Universe. For all these applications it is
important to have an understanding of the kinematics of the LMC and
the SMC, as well the kinematics (i.e., orbit) of their center of mass
with respect to the Milky Way and with respect to each other. These
topics form the subject of the present review. Other related topics,
such as the more general aspects of the structure of the LMC and SMC,
the nature of the LMC bar, the possible presence of fore- or
background populations, and the large radii extent of the Clouds are
not discussed here. The nature, origin, and models of the Magellanic
Stream are touched upon only briefly. All these topics are reviewed in
others papers in this volume by, e.g., Harris, Majewski, Besla, Bekki,
and others.

\section{LMC Kinematics}
\label{s:LMCkin}
Kinematical observations for the LMC have been obtained for many
tracers. The kinematics of gas in the LMC has been studied primarily
using HI (e.g., Kim \etall 1998; Olsen \& Massey 2007). Discrete LMC
tracers which have been studied kinematically include star clusters
(e.g., Schommer \etall 1992; Grocholski \etall 2006), planetary
nebulae (Meatheringham \etall 1988), HII regions (Feitzinger,
Schmidt-Kaler \& Isserstedt 1977), red supergiants (Olsen \& Massey
2007), red giant branch (RGB) stars (Zhao \etall 2003; Cole \etall
2005), carbon stars (e.g., van der Marel \etall 2002; Olsen \& Massey
2007) and RR Lyrae stars (Minniti \etall 2003; Borissova \etall
2005). For the majority of tracers, the line-of-sight velocity
dispersion is at least a factor $\sim 2$ smaller than their rotation
velocity. This implies that on the whole the LMC is a (kinematically
cold) disk system.

\subsection{General Expressions}
\label{ss:generalexp}

To understand the kinematics of an LMC tracer population it is
necessary to have a general model for the line-of-sight velocity field
that can be fit to the data. All studies thus far have been based on
the assumption that the mean streaming (i.e., the rotation) in the
disk plane can be approximated to be circular. However, even with this
simplifying assumption it is not straightforward to model the
kinematics of the LMC. Its main body spans more than $20^{\circ}$ on
the sky and one therefore cannot make the usual approximation that
``the sky is flat'' over the area of the galaxy. Spherical
trigonometry must be used, which yields the general expression (van
der Marel \etall 2002; hereafter vdM02):
\begin{eqnarray}
\label{vfield}
   v_{\rm los}(\rho,\Phi) 
         &&= s \, V(R') f \sin i \cos (\Phi-\Theta) 
           + v_{\rm sys} \cos \rho \nonumber \\
         &&+ v_t \sin \rho \cos (\Phi -\Theta_t) 
           + D_0 (di/dt) \sin \rho \sin (\Phi-\Theta) , 
\end{eqnarray}
with 
\begin{equation}
\label{Rfdef}
  R' = D_0 \sin \rho / f , \qquad 
  f        \equiv { {\cos i \cos \rho -
                     \sin i \sin \rho \sin (\Phi-\Theta)} \over 
                    { {[\cos^2 i \cos^2 (\Phi-\Theta) + 
                                 \sin^2 (\Phi-\Theta)]^{1/2}} } } .
\end{equation}
Here, $v_{\rm los}$ is the observed component of the velocity along
the line of sight. The quantities $(\rho,\Phi)$ identify the position
on the sky with respect to the center: $\rho$ is the angular distance
and $\Phi$ is the position angle (measured from North over East). The
kinematical center is at the center of mass (CM) of the galaxy. The
quantities $(v_{\rm sys}, v_t, \Theta_t)$ describe the velocity of the
CM in an inertial frame in which the sun is at rest: $v_{\rm sys}$ is
the systemic velocity along the line of sight, $v_t$ is the transverse
velocity, and $\Theta_t$ is the position angle of the transverse
velocity on the sky. The angles $(i,\Theta)$ describe the direction
from which the plane of the galaxy is viewed: $i$ is the inclination
angle ($i=0$ for a face-on disk), and $\Theta$ is the position angle
of the line of nodes (the intersection of the galaxy plane and the sky
plane). The velocity $V(R')$ is the rotation velocity at cylindrical
radius $R'$ in the disk plane. $D_0$ is the distance to the CM, and
$f$ is a geometrical factor. The quantity $s = \pm 1$ is the `spin
sign' that determines in which of the two possible directions the disk
rotates.

The first term in equation~(\ref{vfield}) corresponds to the internal
rotation of the LMC. The second term is the part of the line-of-sight
velocity of the CM that is seen along the line of sight, and the third
term is the part of the transverse velocity of the CM that is seen
along the line of sight. For a galaxy that spans a small area on the
sky (very small $\rho$), the second term is simply $v_{\rm sys}$ and
the third term is zero. However, the LMC does not have a small angular
extent and the inclusion of the third term is particularly
important. It corresponds to a solid-body rotation component that at
most radii exceeds in amplitude the contribution from the intrinsic
rotation of the LMC disk. The fourth term in equation~(\ref{vfield})
describes the line-of-sight component due to changes in the
inclination of the disk with time, as are expected due to precession
and nutation of the LMC disk plane as it orbits the Milky Way
(Weinberg 2000). This term also corresponds to a solid-body rotation
component.

The general expression in equation~(\ref{vfield}) appears complicated,
but it is possible to gain intuitive insight by considering some
special cases. Along the line of nodes one has that $\sin
(\Phi-\Theta) = 0$ and $\cos (\Phi-\Theta) = \pm 1$, so that
\begin{equation}
\label{vlosalong}
   {\hat v}_{\rm los} ({\rm along}) = 
          \pm [v_{tc} \sin \rho - 
               V(D_0 \tan \rho) \sin i \cos \rho] .
\end{equation}
Here it has been defined that ${\hat v}_{\rm los} \equiv v_{\rm los} -
v_{\rm sys} \cos \rho \approx v_{\rm los} - v_{\rm sys}$. The quantity
$v_{tc} \equiv v_t \cos (\Theta_t - \Theta)$ is the component of the
transverse velocity vector in the plane of the sky that lies along the
line of nodes; similarly, $v_{ts} \equiv v_t \sin (\Theta_t - \Theta)$
is the component perpendicular to the line of nodes. Perpendicular to
the line of nodes one has that $\cos (\Phi-\Theta) = 0$ and $\sin
(\Phi-\Theta) = \pm 1$, and therefore
\begin{equation}
\label{vlosperp}
   {\hat v}_{\rm los} ({\rm perpendicular}) = \pm w_{ts} \sin \rho .
\end{equation}
Here it has been defined that $w_{ts} = v_{ts} + D_0 (di/dt)$. This
implies that perpendicular to the line of nodes ${\hat v}_{\rm los}$
is linearly proportional to $\sin \rho$. By contrast, along the line
of nodes this is true only if $V(R')$ is a linear function of
$R'$. This is not expected to be the case, because galaxies do not
generally have solid-body rotation curves; disk galaxies tend to have
flat rotation curves, at least outside the very center. This implies
that, at least in principle, both the position angle $\Theta$ of the
line of nodes and the quantity $w_{ts}$ are uniquely determined by the
observed velocity field: $\Theta$ is the angle along which the
observed ${\hat v}_{\rm los}$ are best fit by a linear proportionality
with $\sin \rho$, and $w_{ts}$ is the proportionality constant.

\begin{figure}
\null\bigskip
\epsfxsize=0.8\hsize
\centerline{\epsfbox{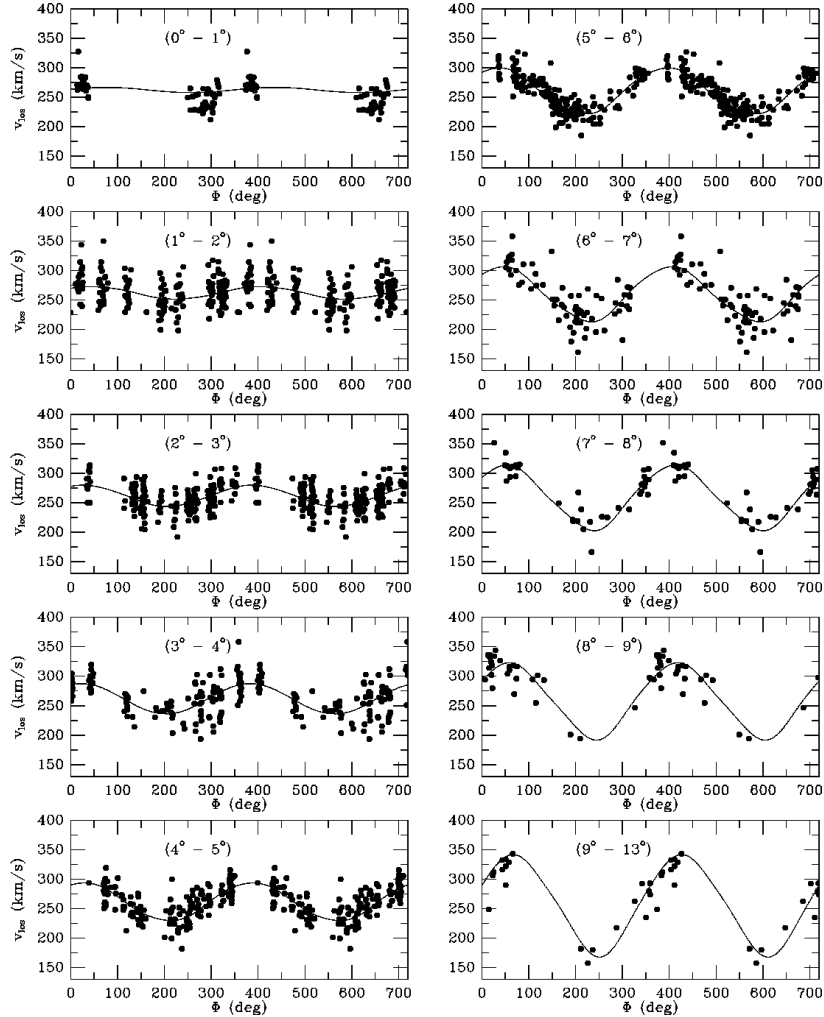}}
\caption{Carbon star line-of-sight velocity data from Kunkel \etall
(1997) and Hardy \etall (unpublished), as a function of position angle
$\Phi$ on the sky. The displayed range of the angle $\Phi$ is
$0^{\circ}$--$720^{\circ}$, so each star is plotted twice. Each panel
corresponds to a different range of angular distances $\rho$ from the
LMC center, as indicated. The curves show the predictions of the
best-fitting circularly-rotating disk model from vdM02.\label{f:fit}}
\end{figure}

\subsection{Carbon Star Kinematics}
\label{ss:carbonstars}

vdM02 were the first to fit the velocity field expression in
equation~(\ref{vfield}) in its most general form to a large sample of
discrete LMC velocities. They modeled the data for 1041 carbon stars,
obtained from the work of Kunkel, Irwin \& Demers (1997) and Hardy,
Schommer \& Suntzeff (unpublished). The combined dataset samples both
the inner and the outer parts of the LMC, although with a
discontinuous distribution in radius and position
angle. Figure~\ref{f:fit} shows the data, with the best model fit
overplotted. Overall, the model provides a good fit to the data.
Olsen \& Massey (2007) recently remodeled the same carbon star data
(for which they obtained a similar fit as vdM02), as well as a large
sample of red supergiant stars.

%%% FIGURE 1 %%%

%%% END OF FIGURE 1 %%%

\subsection{Viewing Angles and Ellipticity}
\label{ss:viewing}

The LMC inclination cannot be determined kinematically, but the
line-of-nodes position angle can. vdM02 obtained $\Theta =
129.9^{\circ} \pm 6.0^{\circ}$ for carbon stars, whereas Olsen \&
Massey (2007) obtained $\Theta = 145.3^{\circ}$ for red supergiants.

A more robust way to determine the LMC viewing angles is to use
geometrical considerations, rather than kinematical ones (since this
avoids the assumption that the orbits are circular). For an inclined
disk, one side will be closer to us than the other. Tracers on that
one side will appear brighter than similar tracers on the other side.
To lowest order, the difference in magnitude between a tracer at the
galaxy center and a similar tracer at a position $(\rho,\Phi)$ in the
disk (as defined in Section~\ref{ss:generalexp}) is
\begin{equation}
\mu = 
   \Big ( { {5\,\pi} \over {180\ln 10} } \Big ) \> 
     \rho \tan i \sin(\Phi-\Theta) ,
\label{dmagtaylor}
\end{equation}
where the angular distance $\rho$ is expressed in degrees. The
constant in the equation is $(5\pi) / (180\ln 10) = 0.038$
magnitudes. Hence, when following a circle on the sky around the
galaxy center one expects a sinusoidal variation in the magnitudes of
tracers. The amplitude and phase of the variation yield estimates of
the viewing angles $(i,\Theta)$.

Van der Marel \& Cioni (2001) used a polar grid on the sky to divide
the LMC area into several rings, each consisting of a number of
azimuthal segments. The data from the DENIS and 2MASS surveys were
used for each segment to construct near-IR color-magnitude diagrams
(CMDs). For each segment both the modal magnitude of carbon stars
(selected by color) and the magnitude of the RGB tip (TRGB) were
determined. This revealed the expected sinusoidal variations at high
significance, implying viewing angles $i = 34.7^{\circ} \pm
6.2^{\circ}$ and $\Theta = 122.5^{\circ} \pm 8.3^{\circ}$. There is an
observed drift in the center of the LMC isophotes at large radii which
is consistent with this result, when interpreted as a result of
viewing perspective (van der Marel 2001). Also, Grocholski \etall
(2006) found that the red clump distances to LMC star clusters are
consistent with a disk-like configuration with these same viewing
angles.

The aforementioned analyses are sensitive primarily to the structure
of the outer parts of the LMC. Several other studies of the viewing
angles have focused mostly on the region of the bar, which samples
only the central few degrees. Nikolaev \etall (2004) analyzed a sample
of more than 2000 Cepheids with lightcurves from MACHO data and
obtained $i = 30.7^{\circ} \pm 1.1^{\circ}$ and $\Theta =
151.0^{\circ} \pm 2.4^{\circ}$. Persson \etall (2004) obtained $i =
27^{\circ} \pm 6^{\circ}$ and $\Theta = 127^{\circ} \pm 10^{\circ}$
from a much smaller sample of 92 Cepheids. Olsen \& Salyk (2002)
obtained $i = 35.8^{\circ} \pm 2.4^{\circ}$ and $\Theta = 145^{\circ}
\pm 4^{\circ}$ from an analysis of variations in the magnitude of the
red clump.

In summary, all studies agree that $i$ is approximately in the range
$30^{\circ}$--$35^{\circ}$, whereas $\Theta$ appears to be in the
range $120^{\circ}$--$150^{\circ}$. The variations between results
from different studies may be due to a combination of systematic
errors, spatial variations in the viewing angles (warps and twists of
the disk plane; van der Marel \& Cioni 2001; Olsen \& Salyk 2002;
Subramaniam 2003; Nikolaev \etall 2004) combined with differences in
spatial sampling between studies, contamination by possible out of
plane structures, and differences between different tracer
populations.

The LMC consists of an outer body that appears elliptical in
projection on the sky, with a pronounced, off-center bar. The
appearance in the optical wavelength regime is dominated by regions of
strong star formation, and patchy dust absorption. However, when only
RGB and carbon stars are selected from near-IR surveys such as 2MASS,
the appearance of the LMC morphology is actually quite regular and
smooth, apart from the central bar. Van der Marel (2001) found that at
radii $r \gta 4^{\circ}$ the contour shapes converge to an
approximately constant position angle ${\rm PA}_{\rm maj} =
189.3^{\circ} \pm 1.4^{\circ}$ and ellipticity $\epsilon = 0.199 \pm
0.008$. A disk that is intrinsically circular will appear elliptical
in projection on the sky, with the major axis position angle ${\rm
PA}_{\rm maj}$ of the projected body equal to the line-of-nodes
position angle $\Theta$. The fact that for the LMC $\Theta \not= {\rm
PA}_{\rm maj}$ implies that the LMC cannot be intrinsically
circular. When the LMC viewing angles are used to deproject the
observed morphology, this yields an in-plane ellipticity $\epsilon$ in
the range $\sim 0.2$--$0.3$. This is larger than typical for disk
galaxies, and is probably due to tidal interactions with either the
SMC or the Milky Way.

\subsection{Transverse Motion and Kinematical Distance}
\label{ss:vtrans}

As discussed in Section~\ref{ss:generalexp}, the line-of-sight
velocity field constrains the value of $w_{ts} = v_{ts} + D_0
(di/dt)$. The carbon star analysis in vdM02 yields $w_{ts} = -402.9
\pm 13.0 \kms$. With the assumptions of a known LMC distance $D_0 =
50.1 \pm 2.5 \kpc$ (based on the distance modulus $m - M = 18.50 \pm
0.10$ adopted by Freedman \etall 2001 on the basis of a review of all
published work) and a constant inclination angle with time ($di/dt =
0$) this yields an estimate of one component of the LMC transverse
velocity. Some weaker constraints can also be obtained for the second
component. The resulting region in LMC transverse velocity space
implied by the carbon star velocity field is shown in Figure~8 of
vdM02. This region is entirely consistent with the Hubble Space
Telescope (HST) proper motion determination discussed in
Section~\ref{ss:PM} below, and therefore provides an important
consistency check on the latter. Alternatively, one can use the HST
proper motion determination with the measured $w_{ts}$ and the
assumption that $di/dt = 0$ to obtain a kinematic distance estimate
for the LMC. This yields $m-M = 18.57 \pm 0.11$, quite consistent with
the Freedman \etall (2001) value.

Previous proper motion estimates for the LMC were lower than the
current HST measurements. This introduced artifacts in previous
analyses of the internal LMC velocity field (which ultimately depends
on subtraction of the $v_t$ term in eq.~[\ref{vfield}] from the
observed line-of-sight velocity field). For example, the HI velocity
field of the LMC presented by Kim \etall (1998) showed a pronounced
S-shape in the zero-velocity contour. Olsen \& Massey recently showed
that this S-shape straightens out when the LMC HST proper motion
measurement is used instead. So this too provides an independent
consistency check on the validity of the HST proper motion
measurement.

\subsection{Rotation Curve and Mass}
\label{ss:rotcurve}

The rotation curve of the LMC rises approximately linearly to $R'
\approx 4 \kpc$, and stays roughly flat at a value $V_{\rm rot}$
beyond that. The carbon star analysis of vdM02 with the HST proper
motion measurement yields $V_{\rm rot} = 61 \kms$. By contrast, for HI
one obtains $V_{\rm rot} = 80 \kms$ and for red supergiants $V_{\rm
rot} = 107 \kms$ (Olsen \& Massey 2007). The random errors on these
numbers are only a few km/s in each case, due to the large numbers of
independent velocity samples. Piatek \etall (2008; hereafter P08)
recently argued for an even higher $V_{\rm rot} = 120 \pm 15 \kms$
based on rotation measurements in the plane of the sky (based on the
same proper motion observations discussed in Section~\ref{ss:PM}
below). All of these measurements are influenced by uncertainties in
the LMC inclination.  However, the uncertainties have different sign
for the $V_{\rm rot}$ values inferred from line-of-sight velocities
than and for those inferred from proper motions. The $V_{\rm rot}$
measurement of P08 becomes more consistent with the line-of-sight
measurements if the inclination is lower than the canonical values
quoted in Section~\ref{ss:viewing}. The $V_{\rm rot}$ estimates from
line-of-sight velocities all have an additional uncertainty due to
uncertainties in the LMC transverse motion. In the end, all $V_{\rm
rot}$ estimates therefore have a systematic error of $\sim
10\kms$.\looseness=-2

The differences between the $V_{\rm rot}$ estimates for various
tracers are significant, and cannot be attributed to either random or
systematic errors. The fact that the HI gas rotates faster than the
carbon stars can probably be largely explained as a result of
asymmetric drift, with the velocity dispersion of the carbon stars
being higher than that of the HI gas (see Section~\ref{ss:disp}
below). The difference between the rotation velocities of HI and red
supergiants is more puzzling, and may point towards non-equilibrium
dynamics. There are in fact clear disturbances in the kinematics of
the various tracers (Olsen \& Massey 2007). Moreover, the dynamical
center of the HI is offset by $\sim 1 \kpc$ from the dynamical and
photometric center of the stars (see e.g. Cole \etall 2005 for a
visual representation of the various relevant centroids of the
LMC). All this complicates the inference of the underlying circular
velocity of the gravitational potential.

If we use the $V_{\rm rot}$ of the HI as a proxy for the circular
velocity, and use the fact that the carbon star rotation curve remains
flat out to the outermost datapoint at $\sim 9$ kpc, then the implied
LMC mass is $M_{\rm LMC} (9 \kpc) = (1.3 \pm 0.3) \times 10^{10}
\Msun$. The mass will continue to rise linearly beyond that radius for
as long as the rotation curve remains flat. By contrast, the total
stellar mass of the LMC disk is $\sim 2.7 \times 10^9 \Msun$ and the
mass of the neutral gas in the LMC is $\sim 0.5 \times 10^9 \Msun$
(Kim \etall 1998). The combined mass of the visible material in the LMC
is therefore insufficient to explain the dynamically inferred mass,
and the LMC must be embedded in a dark halo.

\subsection{Velocity Dispersion and Vertical Structure}
\label{ss:disp}

As in the Milky Way, younger populations have a smaller velocity
dispersion (and hence a smaller scale height) than older
populations. Measurements, in order of increasing dispersion, include:
$\sim 9 \kms$ for red supergiants (Olsen \& Massey 2007); $\sim 16
\kms$ for HI gas (Kim \etall 1998); $\sim 20 \kms$ for carbon stars
(vdM02); $\sim 25 \kms$ for RGB stars (Zhao \etall 2003; Cole \etall
2005); $\sim 30 \kms$ for star clusters (Schommer \etall 1992;
Grocholski \etall 2006); $\sim 33 \kms$ for old long-period variables
(Bessell, Freeman \& Wood 1986); $\sim 40 \kms$ for the lowest
metallicity red giant branch stars with [Fe/H] $< -1.15$ (Cole \etall
2005); and $\sim 50 \kms$ for RR Lyrae stars (Minniti \etall 2003;
Borissova \etall 2005).

For the stars with the highest dispersions it has been suggested that
they may form a halo distribution, and not be part of the LMC disk. On
the other hand, this remains unclear, since the kinematics of these
stars have typically been observed only in the central region of the
LMC. Therefore, it is not known whether the rotation properties of
these populations are consistent with being a separate halo
component. In fact, the surface density distribution of the LMC RR
Lyrae stars is well fit by an exponential with the same scale length
as inferred for other tracers known to reside in the disk (Alves
2004). Either way, the vertical extent of all LMC populations is
certainly significant. For example, even the (intermediate-age) carbon
stars only have $V/\sigma \approx 3$. For comparison, the thin disk of
the Milky Way has $V /\sigma \approx 9.8$ and its thick disk has
$V/\sigma \approx 3.9$.

The velocity residuals with respect to a rotating disk model do not
necessarily follow a Gaussian distribution. Although Zhao \etall
(2003) did not find large deviations from a Gaussian for RGB stars,
Graff \etall (2000) found that the carbon star residuals are better
fit by a sum of Gaussians. More recently, Olsen \& Massey (2007)
showed that some fraction of both carbon stars and red supergiants
have peculiar kinematics that suggest an association with tidally
disturbed features previously identified in HI.\looseness=-2

\section{SMC Kinematics}
\label{s:SMCkin}

The SMC structure and kinematics are less well studied and understood
than those of the LMC. The morphological appearance in blue optical
light is patchy and irregular. Kinematical observations of HI and
young stars reveal ordered rotation that indicates that these tracers
may reside in a disk. However, detailed velocity field fits using
equation~(\ref{vfield}) have not been attempted. Stanimirovic \etall
(2004) found that the HI rotation curve in the SMC rises almost
linearly to $V_{\rm rot} \approx 50 \kms$ at the outermost datapoint
($\sim 3.5 \kpc$), with no signs of flattening. The implied dynamical
mass inside this radius is $2.4 \times 10^9 \Msun$. By contrast, the
total stellar mass of the SMC is $\sim 3.1 \times 10^8 \Msun$ and the
mass of the neutral gas is $5.6 \times 10^8 \Msun$. The combined mass
of the visible material in the SMC is therefore insufficient to
explain the dynamically inferred mass, and the SMC must be embedded in
a dark halo.\footnote{These values are based on the analysis in
Stanimirovic \etall (2004), although those authors do not draw the
same conclusion.}

Evans \& Howarth (2008) obtained velocities for 2045 young (O, B, A)
stars in the SMC, and found a velocity gradient of similar slope as
seen in the HI gas. Surprisingly though, they find a position angle
for the line of maximum velocity gradient that is quite different, and
almost orthogonal to that seen in the HI. This may be an artifact of
the different spatial coverage of the two studies (Evans \& Howarth
did not observe in the North-East region where the HI velocities are
largest), since it would be difficult to find a physical explanation
for a significant difference in kinematics between HI gas and young
stars.\looseness=-2

When the old red stars that trace most of the stellar mass are
isolated using CMDs, the morphological appearance of the SMC is more
spheroidal (Zaritsky \etall 2000, 2002 Cioni, Habing \& Israel 2000b;
Maragoudaki \etall 2001). Harris \& Zaritsky (2006) studied the
kinematics of 2046 RGB stars and inferred a velocity dispersion
$\sigma = 27.5 \pm 0.5 \kms$. This is similar to the dispersion of the
young stars observed by Evans \& Howarth (2008), but unlike the young
stars, the older RGB stars do not show much rotation. Their low
$V_{\rm rot}/\sigma$ is consistent with what is typical for dE and
dSph galaxies. Hence, the SMC may be more akin to those galaxy types
than to other more irregular systems.

Studies of the distances of individual tracers in the SMC have shown
it to be much more vertically extended than would be expected for a
disk galaxy. Crowl \etall (2001) mapped the distances of star clusters
using red clump magnitudes.  They argued that the SMC has axial ratios
of 1:2:4, and is viewed almost pole on. While different authors have
found a range of other axial ratios using different types of tracers,
most authors agree that the SMC has a considerable line-of-sight
depth.

\section{Orbital History of the Magellanic Clouds}
\label{s:orbhist}

To understand the history of the Magellanic System and the origin of
the Magellanic Stream it is important to know the orbit of the
Magellanic Clouds around the Milky Way. This requires for each Cloud
knowledge of all three of the velocity components of the center of
mass. Line-of-sight velocities can be accurately determined from the
Doppler velocities of tracers. However, determination of the velocity
in the plane of the sky through proper motions is much more difficult.
This was the primary obstacle for a long time, but recent
breakthroughs have now yielded considerable progress.

\subsection{Proper Motions}
\label{ss:PM}

Previous attempts at measuring the proper motions (PMs) of the LMC and
SMC from the ground or with Hipparcos were reviewed in vdM02. However,
this earlier work has now been largely superseded by the studies
performed with the ACS/HRC on HST by Kallivayalil \etall (2006a,b;
hereafter K06a,b). Two epochs of data were obtained with a $\sim 2$
year time baseline for 21 fields in the LMC and 5 fields in the SMC,
all centered on background quasars identified from the MACHO database
(Geha \etall 2003). PMs were obtained for each field by measuring the
average shift of the stars with respect to the background quasar. Upon
correction for the orientation and rotation of the LMC disk, each
field yields an independent estimate of the center-of-mass PM. The
average for the different fields yields the final estimate, while the
RMS among the results from the $N$ different fields yields the PM
error RMS$/\sqrt{N}$.

K06a,b obtained for the PMs in the West and East directions that
\begin{eqnarray}
\label{K06}
   \mu_W &=& -2.03 \pm 0.08 \masyr , \qquad 
             \mu_N = 0.44 \pm 0.05 \masyr \qquad (LMC) \nonumber \\
   \mu_W &=& -1.16 \pm 0.18 \masyr , \qquad 
             \mu_N = -1.17 \pm 0.18 \masyr \qquad (SMC) .
\end{eqnarray}
The same data were reanalyzed more recently by P08. Using an
independent analysis with different software and point spread
function models they obtained that
\begin{eqnarray}
\label{P08}
   \mu_W &=& -1.96 \pm 0.04 \masyr , \qquad 
             \mu_N = 0.44 \pm 0.04 \masyr \qquad (LMC) \nonumber \\
   \mu_W &=& -0.75 \pm 0.06 \masyr , \qquad 
             \mu_N = -1.25 \pm 0.06 \masyr \qquad (SMC) .
\end{eqnarray}
P08 made magnitude-dependent corrections for small charge transfer
inefficiency effects. By contrast, K06a,b assumed that these effects
(always along the detector y axis) average to zero over all fields
because of the random telescope orientations used for different
fields. The fact the the results from these two studies are generally
in good agreement for the LMC confirms the validity of this
assumption. However, the explicit correction applied by P08 does yield
better agreement between different fields, and therefore smaller
errorbars.  For the SMC, the results for the individual fields are in
good agreement between the studies. However, the studies used
different methods for weighted averaging of the fields, with K06a,b
being more conservative and allowing for potential unknown systematic
effects. This produces both larger errorbars than in P08, as well as a
significant difference in $\mu_W$.\looseness=-2

Transformation of the PMs to a space velocity in $\kms$ requires
knowledge of the distance $D_0$. For the LMC, $D_0 \approx 50.1 \kpc$,
so that $1 \masyr$ corresponds to $238 \kms$. For the SMC, $D_0
\approx 61.6 \kpc$, so that $1 \masyr$ corresponds to $293
\kms$. After transformation of the PM to $\kms$, it can be combined
with the observed center-of-mass line-of-sight velocity to obtain the
full three-dimensional velocity vector. For the LMC, $v_{\rm sys} =
262.2 \pm 3.4 \kms$ (vdM02); and for the SMC, $v_{\rm sys} = 146 \pm
0.6 \kms$ (Harris \& Zaritsky 2006). The resulting vectors can be
corrected for the solar reflex motion and transformed to the
Galactocentric rest-frame as described in vdM02. For the LMC this
yields that the motion has a radial component of $V_{\rm rad} = 89 \pm
4 \kms$ pointing away from the Galactic center, and a tangential
component of $V_{\rm tan} = 367 \pm 18 \kms$.\footnote{These
velocities are based on the K06a PM values. However, use of the P08 PM
values would yield similar velocities that would not alter the
arguments in the remainder of the text.}

\subsection{Orbit around the Milky Way}
\label{ss:orbMW}

The combination of a small but positive radial velocity and a
tangential velocity that exceeds the circular velocity of the Milky
Way halo implies that the Clouds must be just past pericenter. The
calculation of an actual orbit requires detailed knowledge of the
gravitational potential of the Milky Way dark halo. Past work had
generally assumed that the dark halo can be approximated by a
spherical logarithmic potential.  Estimates of the transverse
velocities of the Clouds based on models of the Magellanic Stream had
suggested that for the LMC $V_{\rm tan} = 287 \kms$ (e.g., Gardiner,
Sawa, \& Fujimoto 1994; Gardiner \& Noguchi 1996). This then yielded
an orbit with an apocenter to pericenter ratio of $\sim 2.6:1$ and an
orbital period of $\sim 1.6 \Gyr$. This orbit was adopted by most
subsequent modeling studies of the Magellanic System. However, the new
HST PM measurements significantly revise this view. The observed
$V_{\rm tan}$ is $80 \pm 16 \kms$ larger than the Gardiner \& Noguchi
(1996) value, and therefore inconsistent with it at $\sim
5\sigma$. The observed value implies a much larger apocenter distance
(in excess of $200 \kpc$) at which the assumption of a logarithmic
potential is not a good assumption.

Motivated by these considerations, Besla \etall (2007) performed a new
study of the Magellanic Cloud orbits using an improved Milky Way
model, combined with the K06a,b HST PMs. The Milky Way model was
chosen similar to that proposed in Klypin \etall (2002). It consists
of disk, bulge, hot gaseous halo, and dark halo components. The dark
halo has a $\Lambda$CDM-motivated NFW potential with adiabatic
contraction. In the fixed Milky Way potential, the orbits of the LMC
and SMC were integrated backwards in time, starting from the current
observed positions and velocities. The extent of the galaxies was
taken into account in the calculation of their mutual gravitational
interaction, and a parameterized prescription was used to account for
dynamical friction. The gravitational influence of M31 can be taken
into account in this formalism as well, but this make little
difference to the results (Kallivayalil 2007; Shattow \& Loeb 2008).

The most favored Milky Way model presented by Klypin \etall (2002) has
a total mass $M = 10^{12} \Msun$. In this model, the escape velocity
at $50 \kpc$ is $\sim 380 \kms$. This is very similar to the observed
$V_{\rm tan}$ of the LMC, and as a result, the inferred orbit is
approximately parabolic, with no previous pericenter passage. In other
words, the Magellanic Clouds are passing by the Milky Way now for the
first time. To obtain an orbit that is significantly bound, $\mu_W$
would have to be larger by $\sim +0.3 \masyr$ ($4\sigma$ with the K06a
errorbar, or $7\sigma$ with the P08 errorbar). Alternatively, it is
possible that the Milky Way is more massive (Smith \etall 2007;
Shattow \& Loeb 2008). A mass of $M = 2 \times 10^{12} \Msun$ is more
or less the largest mass consistent with the available observational
constraints (see also the discussion in van der Marel \& Guhathakurta
2008). This would produce a bound orbit. However, with either the
larger Milky Way mass or with the larger $\mu_W$, the orbit would
still be quite different than has been previously assumed in models of
the Magellanic System. There would be only 1 previous pericenter
passage, the apocenter distance would be $400 \kpc$ or more, and the
period would be 6-7 Gyr.\footnote{if this were in fact the case, then
the mass build-up of the Milky Way with time, as in e.g. Wechsler
(2002), would also have to be taken into account for calculation of an
accurate orbit.} Therefore, the new PM results drastically alter our
view of the history of the Magellanic System.

The view that the Magellanic Clouds may be passing by the Milky Way
for the first time may seem revolutionary at first. However, there are
arguments to consider this reasonable. Van den Bergh (2006) pointed
out that the LMC and SMC are unusual in that they are the only
satellites in the Local Group that are both gas rich and located close
to their parent galaxy. He suggested based on this that the Magellanic
Clouds are interlopers that were originally formed in the outer
reaches of the Local Group. Moreover, cosmological simulations show
that: (a) accretion of LMC-sized subhalos by Milky-Way sized halos is
common since $z \sim 1$; and (b) finding long-term satellites with
small pericenter distances around Milky-Way sized halos is rare
(Kazantzidis \etall 2008). Therefore, a scenario in which the
Magellanic Clouds are passing by the Milky Way now for the first time
seems more likely from a purely cosmological perspective than a
scenario in which they have been satellites for many orbital periods
of 1--2 Gyr each.

\subsection{LMC--SMC orbit}
\label{ss:orbLS}

Although the Magellanic Clouds may not bound to the Milky Way, it
would be much more unlikely for the Magellanic Clouds not to be bound
to each other. The likelihood of two satellite galaxies running into
each other by chance is quite low. Also, various properties of the
Clouds (such as their common HI envelope) suggest that they have been
associated with each other for a significant time. The K06a,b PMs
imply a relative velocity between the SMC and LMC of $105 \pm 42
\kms$. Orbit calculations (K06b; Besla \etall 2007) show that the
error bar on this is too large to say with any confidence whether or
not they are indeed bound. However, binary orbits do exist within the
$1\sigma$ error ellipse, so there seems little reason to depart from
this null hypothesis. Indeed, there are allowed orbits that have close
passages between the Clouds at $\sim 0.3 \Gyr$ and $\sim 1.5 \Gyr$ in
the past (Besla et al., these proceedings). These are the time scales
that have been previously associated with the formation of the
Magellanic Bridge and Stream, respectively.

\subsection{Magellanic Stream}
\label{ss:stream}

The Magellanic Stream is discussed in detail in other contributions in
this volume. One important thing to note though in the present
context is that the new insights into the orbit of the Magellanic
Clouds around the Milky Way drastically affect our understanding of
the Magellanic Stream. The Stream does not lie along the projected
path on the sky traced by the LMC and SMC orbits, and the HI velocity
along the Stream is not as steep as that along the orbits (Besla
\etall 2007). This is inconsistent with purely tidal models of the
stream (e.g., Gardiner \& Noguchi 1996; Connors, Kawata \& Gibson
2006).  Moreover, the more limited number of passages through the
Milky Way disk, and the larger radius at which this occurs, imply that
the ram pressure models that have been proposed (e.g., Moore \& Davis
1994; Mastropietro \etall 2005) probably won't work either. It is
therefore essential that models of the Magellanic Stream be revisited,
with an eye towards inclusion of new physics and exploration of new
scenarios (see Besla et al., these proceedings). The recent finding
by Nidever, Majewski \& Burton (2008) that one filament of the
Magellanic Stream, containing more than half its gas mass, can be
traced back to the 30 Doradus star forming region in the LMC is
particularly interesting in this context. This indicates that an
outflow may have created or contributed to the Stream, which has not
been addressed in previous models.

\section{Concluding Remarks}
\label{s:conc}

The kinematics of the LMC are now fairly well understood, with
velocities of thousands of individual tracers of various types having
been fitted in considerable detail with (thick) disk models. Questions
that remain open for further study include the reality and origin of
kinematical differences between different stellar tracer populations,
the differences between the gaseous and stellar kinematics, and the
amount and origin of non-equilibrium features in the kinematics. The
kinematics of the SMC are understood more poorly, but appear generally
consistent with being a spheroidal system of old stars with an
embedded irregular disk of gas and young stars.

The HST PM work has provided the most surprising results in recent
years, with important implications for both the history of the
Magellanic System and the origin of the Magellanic Stream. Of course,
it is natural in discussions about this to wonder about the robustness
of the observational results. It should be noted in this context that
many experimental features and consistency checks are built in that
support the general validity of the HST PM results. These include:
(1) the use of random telescope orientations causes systematic errors
tied to the detector frame to cancel out when averaging over all
fields; (2) the final PM errors are based on the observed scatter
between fields, with no assumptions about the source and nature of the
underlying errors; (3) two groups used different methods to analyze
the same data and obtained consistent results; (4) P08 managed to
measure a PM rotation curve for the LMC that is broadly consistent
with expectation, which would have been impossible if the PM errors
were in reality larger than claimed; (5) the difference between the
LMC and SMC PMs is more or less consistent with expectation for a
binary orbit, which would not generally have been the case if the
measurements suffer from unknown systematics; (6) the LMC PM is
consistent with expectation based on the line-of-sight velocity field
of carbon stars (see Section~\ref{ss:vtrans}); and (7) the LMC PM
leads to an HI velocity field with a straight zero-velocity curve, by
contrast to previously assumed values (see Section~\ref{ss:vtrans}).

One interesting feature in the observational PM results is that with
the P08 PM values, there are no bound LMC-SMC orbits, given their
different $\mu_W$ and smaller error bars for the SMC compared to the
K06b results. However, the SMC PM is significantly less certain that
that for the LMC, due to the smaller number of fields observed with
HST, and the fact that most of them were observed at a similar
telescope orientation (which implies that potential systematic errors
that are fixed in the detector frame do not average out when the
results from different fields are combined). This underscores the need
for additional PM observations. A third epoch of observations for most
fields has already been obtained with HST/WFPC2, and preliminary
analysis supports the validity of the results based on the first two
epochs (Kallivayalil et al., these proceedings). A fourth epoch is
planned with HST/ACS and HST/WFC3 in 2009. With the increased time
baselines and use of multiple different instruments it will be
possible to further reduce random errors and constrain possible
systematic errors. In turn, this will allow new scientific problems to
be addressed, such as the internal proper motion kinematics of the
Clouds, and their rotational parallax distances (the distances
obtained by equating the line-of-sight and proper motion rotation
curves).

%%%%%%%%%%%%%%%%%%%%
% References
%%%%%%%%%%%%%%%%%%%%


\begin{thebibliography}{}

\bibitem[]{Alv04}
Alves, D. R. 2004, ApJ, 601, L151

\bibitem[]{Bes08}
Besla, G., Kallivayalil, N, Hernquist, L., Robertson, B., Cox, T. J.,
van der Marel, R. P., \& Alcock, C. 2007, 668, 949

\bibitem[]{Bes86}
Bessell, M. S., Freeman, K. C., \& Wood, P. R. 1986, ApJ, 310, 710

\bibitem[]{Bor06}
Borissova, J., Minniti, D., Rejkuba, M., \& Alves, D.
2006, A\&A, 460, 459

\bibitem[]{Cio00b}
Cioni, M.-R. L., Habing, H. J., \& Israel, F. P. 2000b, A\&A, 358, L9

\bibitem[]{Col05}
Cole, A. A., Tolstoy, E., Gallagher, J. S., III, \& Smecker-Hane, T. A.
2005, AJ, 129, 1465

\bibitem[]{Con06}
Connors, T. W., Kawata, D., \& Gibson, B. K. 2006, MNRAS, 371, 108

\bibitem[]{Cro01}
Crowl, H. H., Sarajedini, A., Piatti, A. E., Geisler, D., Bica, E., 
Claria, J. J., \& Santos, J. F. C., Jr. 2001, AJ, 122, 220

\bibitem[]{Eva08}
Evans, C. J., \& Howarth, I. D. 2008, MNRAS, 386, 826

\bibitem[]{Fei77}
Feitzinger, J. V., Schmidt-Kaler, T., \& Isserstedt, J. 1977, A\&A, 57, 265

\bibitem[]{Fre01}
Freedman, W. L., et al. 2001, ApJ, 553, 47

\bibitem[]{Gar96}
Gardiner, L. T., \& Noguchi, 1996, MNRAS, 278, 191

\bibitem[]{Gar94}
Gardiner, L. T., Sawa, T., \& Fujimoto, M. 1994, MNRAS, 266, 567

\bibitem[]{Geh03}
Geha, M., et al. 2003, AJ, 125, 1

\bibitem[]{Gra00}
Graff, D. S., Gould, A. P., Suntzeff, N. B., Schommer, R. A., \&
Hardy, E. 2000, ApJ, 540, 211

\bibitem[]{Gro06}
Grocholski, A. J., Cole, A. A., Sarajedini, A., Geisler, D., \& Smith, V. V.
2006, AJ, 132, 1630

\bibitem[]{Har06}
Harris, J., \& Zaritsky, D. 2006, AJ, 131, 2514

\bibitem[]{Kal06a}
Kallivayalil, N., van der Marel, R. P., Alcock, C., Axelrod, T., Cook, K. H.,
Drake, A. J., \& Geha, M. 2006a, ApJ, 638, 772 (K06a)

\bibitem[]{Kal06b}
Kallivayalil, N., van der Marel, R. P., \& Alcock, C. 2006b,
ApJ, 652, 1213 (K06b)

\bibitem[]{Kal07}
Kallivayalil, N. 2007, PhD thesis, Harvard University 

\bibitem[]{Kaz08}
Kazantzidis, S., Zentner, A. R., \& Bullock, J. S. 2008, ApJ, in press
[arXiv:0807.2863]

\bibitem[]{Kim98}
Kim, S., Staveley-Smith, L., Dopita, M. A., Freeman, K. C., Sault, R. J., 
Kesteven M. J., \& McConnell, D. 1998, ApJ, 503, 674

\bibitem[]{Kly02}
Klypin, A., Zhao., H. S., \& Somerville, R. S. 2002, 573, 597 

\bibitem[]{Kun97}
Kunkel, W. E., Irwin, M. J., \& Demers, S. 1997, A\&AS, 122, 463

\bibitem[]{Mar01}
Maragoudaki, F., Kontizas, M., Morgan, D. H., Kontizas, E., Dapergolas, A.,
\& Livanou, E. 2001, A\&A, 379, 864

\bibitem[]{Mas05}
Mastropietro, C., Moore, B., Mayer, L., Wadsley, J., \& Stadel, J. 2005,
363, 509

\bibitem[]{Mae88}
Meatheringham, S. J., Dopita, M. A., Ford, H. C., \& Webster, B. L. 1988,
ApJ, 327, 651

\bibitem[]{Min03}
Minniti, D., Borissova, J., Rejkuba, M., Alves, D. R., Cook, K. H., \& Freeman, 
K. C. 2003, Science, 301, 1508

\bibitem[]{Moo94}
Moore, B., \& Davis, M. 1994, MNRAS, 270, 209

\bibitem[]{Nid08}
Nidever, D. L., Majewski, S. R., \& Burton, W. B. 2008, ApJ, 679, 432

\bibitem[]{Nik04}
Nikolaev, S., Drake, A.J., Keller, S. C., Cook, K. H., Dalal, N., Griest, K., 
Welch, D. L., \& Kanbur, S. M. 2004, 601, 260

\bibitem[]{Ols02}
Olsen, K. A. G., \& Salyk, C. 2002, AJ, 124, 2045

\bibitem[]{Ols07}
Olsen, K. A. G., \& Massey, P. 2007, ApJ, 656, L61O

\bibitem[]{Per04}
Persson, S. E., Madore, B. F., Krzeminski, W., Freedman, W. L., Roth, M., \&
Murphy, D. C. 2004, AJ, 128, 2239

\bibitem[]{Pia08} Piatek, S., Pryor, C., \& Olszewski, E. W. 2008,
ApJ, 135, 1024 (P08)

\bibitem[]{Scho92}
Schommer, R. A., Suntzeff, N. B., Olszewski, E. W., \& Harris, H. C.
1992, AJ, 103, 447

\bibitem[]{Sha08}
Shattow, G., \& Loeb, A. 2008, MNRAS, submitted [arXiv:0808.0104]

\bibitem[]{Smi07}
Smith, M. C., et al. 2007, MNRAS, 379, 755
 
\bibitem[]{Sta04}
Stanimirovic, S., Staveley-Smith, L., \& Jones, P. 2004, ApJ, 604, 176 

\bibitem[]{Sub03}
Subramaniam, A. 2003, ApJ, 598, L19

\bibitem[]{vdB06}
van den Bergh, S. 2006, AJ, 132, 1571

\bibitem[]{vdM01a}
van der Marel, R. P., \& Cioni, M.-R. 2001, AJ, 122, 1807

\bibitem[]{vdM01b}
van der Marel, R. P. 2001, AJ, 122, 1827

\bibitem[]{vdM02}
van der Marel, R. P., Alves, D. R., Hardy, E., \& Suntzeff, N. B. 
2002, AJ, 124, 2639 (vdM02)

\bibitem[]{vdM08}
van der Marel, R. P., \& Guhathakurta, P. 2008, ApJ, 678, 187

\bibitem[]{Wedc02}
Wechsler, R. H., Bullock, J. S., Primack, J. R., Kravtsov, A. V., \& 
Dekel, A. 2002, ApJ, 568, 52

\bibitem[]{Wei00b}
Weinberg, M. D. 2000, ApJ, 532, 922

\bibitem[]{Zar00}
Zaritsky, D., Harris, J., Grebel, E. K., \& Thompson, I. B. 2000, ApJ, 534, 
L53

\bibitem[]{Zar02}
Zaritsky, D., Harris, J., Thompson, I. B., Grebel, E. K., \& Massey, P. 2002,
AJ, 123, 855

\bibitem[]{Zha03}
Zhao, H., Ibata, R. A., Lewis, G. F., \& Irwin, M. J. 2003, MNRAS, 339, 701

\end{thebibliography}
\end{document}